\documentclass[12pt]{article}
\usepackage{bbm}
\usepackage{amsfonts}
 \usepackage{amsmath}

\usepackage{mathrsfs}
\title{Algebro-geometric Constructions to the Dym-type Hierarchy}
\date{}

\author{\leftline{\hspace{3 cm}Lihua Wu$^{1}$, Guoliang He$^{2}$, and Xianguo Geng$^{3}$}\\
\leftline{\hspace{0.6 cm}{\small{ $^{1}$ Department of Mathematics, Huaqiao University,  Quanzhou 362021, P. R. China}}}\\
\leftline{\hspace{0.6 cm}{\small{ $^{2}$ Department of Mathematics and Information Science, Zhengzhou University of Light }}}\\
\leftline{\hspace{0.9 cm}{\small{  Industry, Zhengzhou 450002, China}}}\\
\leftline{\hspace{0.6 cm}{\small{ $^{3}$ School of Mathematics and
Statistics, Zhengzhou University,  Zhengzhou 450001, P. R.
China}}}\\ \\ \leftline{\hspace{0.6 cm}{\small{ {\sl Correspondence
to be sent to: wulihua@hqu.edu.cn}}}}}

\begin{document}

\maketitle {}

 Resorting to the characteristic polynomial of Lax matrix for the
Dym-type hierarchy, we define a trigonal curve, on which
appropriate vector-valued Baker-Akhiezer function and meromorphic function are
introduced. Based on the theory of trigonal curve and three kinds of Abelian differentials, we obtain the explicit
Riemann theta function representations of the meromorphic function, from
which we get the algebro-geometric constructions for the entire
Dym-type hierarchy.

\section{Introduction }\setcounter{equation}{0}
As is well-known, there exist several methods to study the
algebro-geometric solutions of soliton equations and from which many soliton equations associated with
$2\times 2$ matrix spectral problems were discussed (\cite{alberfedorov:2001, caowugeng:1999,Dmitrieva:1993,dubrovin:1975,gengdaicao:2003,gengwucao:1999,gengxue:2010,gesztesyholden:2003,gesztesyratneseelan:1998,its,krichever:1976,maablowitz:1981,matveevsmirnov:1993,miller:1995,novikov:1974,Novikov:1999,previato:1985,qiao:2001,smirnov:1986,
 zengma:1999,zhouma:2000} and references therein). However, the research of soliton equations associated with $3\times 3$ matrix spectral problems is very few, which
is also much more difficult and complicated for the underlying algebraic curve is trigonal curve. It should be pointed out that this trigonal curve considerably complicates the analysis and hence makes it a rather challenging problem. More recently, according to a unified framework
\cite{dickson:1999}, algebro-geometric solutions for a lot of soliton
hierarchies associated with $3\times 3$ matrix spectral problems
have been discussed, such as modified Boussinesq hierarchy \cite{gengwuhe:2011},
Kaup-Kuperschmidt hierarchy \cite{gengwuhe:2013}, three-wave resonant interaction
hierarchy \cite{hegengwu:2014}, and others \cite{gengzhaidai:2014,wugengzhang:2015}.

 The Dym equation
 \begin{equation}
\begin{array}{lcl}
u_{t}=(u^{-\frac{1}{2}})_{xxx},
\end{array}
\end{equation}
was first discovered by Harry Dym \cite{Kruskal:1975} and rediscovered by Li \cite{Li:1982} and Sabatier \cite{Sabatier:1979}.
It was shown that the Dym equation possesses many properties typical for integrable systems (see \cite{CalogeroDegasperis:1982, Magri:1978,WadatiIchikawaShimizu:1980} and references therein).  Moreover, the algebro-geometric solution of
Dym equation was also discussed in \cite{Dmitrieva:1993,Novikov:1999}. Meanwhile, the
integrable extensions of Dym equation attract much attention of many researchers \cite{AntonowiczFordy:1998,KonopelchenkoDubrovsky:1984,Ma:2010, Popowicz:2003}.

By considering a $3\times 3$ matrix spectral problem,
Prof. Geng \cite{geng:1994} derived a hierarchy of Dym-type equations and
discussed its nonlinearization. The first nontrivial member in the
hierarchy is Dym-type equation
\begin{equation}
\begin{array}{lcl}
v_{t}=-\frac{1}{3}(v^{-\frac{2}{3}})_{xxxxx}.
\end{array}
\end{equation}
The principal aim of the present paper is to study algebro-geometric
constructions of the Dym-type flows. With the aid of the three
kinds of Abelian differentials and asymptotic expansions, we arrive
at the Riemann theta function representations of the meromorphic function,
and solutions for the entire Dym-type hierarchy. In this process, an explicit expression of
the third kind of Abelian differential proposed by us is of great importance.

The outline of the present paper is as follows. In section 2, based
on the Lenard recursion equations and the zero-curvature equation,
we deduce the Dym-type hierarchy. In section 3, we define the vector-valued
Baker-Akhiezer function and the associated meromorphic function, from which a
trigonal curve ${\mathcal K}_{m-1}$ of arithmetic genus $m-1$ is
introduced with the help of the characteristic polynomial of Lax
matrix for the Dym-type hierarchy. It is shown that the Dym-type hierarchy is decomposed into a system
of Dubrovin-type equations. In section 4, by
introducing three kinds of Abelian differentials, especially the
explicit third kind, we present the Riemann theta function
representations of the meromorphic function, and in particular, that
of the potential $v$ for the entire Dym-type hierarchy.

\section{Dym-type Hierarchy }\setcounter{equation}{0}
In this section, we follow the Geng \cite{geng:1994} and derive the Dym-type hierarchy
associated with the $3\times3$ matrix spectral problem
\begin{equation}
\begin{array}{rcl}
\psi_x=U\psi,\quad \psi=\left(\begin{matrix}\psi_1\cr \psi_2\cr
\psi_3\end{matrix}\right), \quad U=\left(\begin{matrix}0&1&0\cr
0&0&1\cr \lambda v&0&0\end{matrix}\right), \end{array}
\end{equation}
where $v (\neq 0)$ is a potential and $\lambda$ is a constant spectral
parameter. To this end, we introduce two sets of Lenard recursion
equations
\begin{equation}
\begin{array}{rcl}
Kg_{j-1}=Jg_j,\quad j\geq 0,
\end{array}
\end{equation}
\begin{equation}
\begin{array}{rcl}
K\hat g_{j-1}=J\hat g_j,\quad j\geq 0,
\end{array}
\end{equation}
with two starting points
\begin{equation}
\begin{array}{rcl}
 g_{-1}=\left(\begin{matrix}v^{-\frac{2}{3}}\cr
0\end{matrix}\right),\quad \hat
g_{-1}=\left(\begin{matrix}-\frac{1}{3}v^{-1}(v^{-\frac{1}{3}})_{xx}+\frac{1}{6}v^{-\frac{2}{3}}[({v^{-\frac{1}{3}})_x}]^2\cr
v^{-\frac{1}{3}}\end{matrix}\right),
\end{array}
\end{equation}
 and two operators are defined as
$$K=\left(\begin{matrix}
 -\frac{1}{3}\partial^5&0  \cr
 2\partial v+v\partial &\partial^3\cr
\end{matrix}\right),\quad
J=\left(\begin{matrix}
 0&-(\partial v+2v\partial)\cr
 2\partial v+v\partial &\partial^3\cr\end{matrix}\right).$$
It is easy to see that
$${\rm ker}J=\{\alpha_0g_{-1}+{\beta}_0{\hat g}_{-1}\ |\ \forall\alpha_0,{\beta}_0\in\mathbb{R}\}.$$
In order to generate a hierarchy of evolution equations associated
with the spectral problem (2.1), we solve the stationary
zero-curvature equation
\begin{equation}
\begin{array}{rcl}
V_x-[U,V]=0,\ \ V=\lambda\left(\begin{matrix}V_{11}& V_{12}&
V_{13}\cr V_{21}&V_{22}&V_{23}\cr V_{31}&
V_{32}&V_{33}\end{matrix}\right),
\end{array}
\end{equation}
which is equivalent to
\begin{equation}
\begin{array}{rcl}
&&V_{11,x}+\lambda vV_{13}-V_{21}=0,\\
&&V_{12,x}+V_{11}-V_{22}=0,\\
&&V_{13,x}+V_{12}-V_{23}=0,\\
&&V_{21,x}+\lambda v V_{23}-V_{31}=0,\\
&&V_{22,x}+V_{21}-V_{32}=0,\\
&&V_{23,x}+V_{22}-V_{33}=0,\\
&&V_{31,x}-\lambda v(V_{11}-V_{33})=0,\\
&&V_{32,x}-\lambda vV_{12}+V_{31}=0,\\
&&V_{33,x}-\lambda vV_{13}+V_{32}=0,\\
\end{array}
\end{equation}
where each entry $V_{ij}=V_{ij}(a,b)$ is a Laurent expansion in
$\lambda$:
\begin{equation}
\begin{array}{lcl}
V_{11}=-\frac{1}{3}\partial^2a-\lambda\partial b,\quad
 V_{12}=\partial a+\lambda b,\quad V_{13}=-2a,\\
V_{21}=-\frac{1}{3}\partial^3a-\lambda\partial^2 b-2\lambda
va,\ \  V_{22}=\frac{2}{3}\partial^2a,\ \
V_{23}=- \partial a+\lambda b, \\
V_{31}=-\frac{1}{3}\partial^4a+\lambda^2vb, \ \
V_{32}=\frac{1}{3}\partial^3a-\lambda\partial^2 b-2\lambda va,\quad
V_{33}=-\frac{1}{3}\partial^2a+\lambda\partial b.
\end{array}
\end{equation}
 Substituting (2.7) into (2.6) and expanding the functions $a$ and $b$ into
the Laurent series in $\lambda$
\begin{equation}
\begin{array}{rcl}
a=\sum\limits_{j\geq 0}a_{j-1}\lambda^{-2j},\quad
b=\sum\limits_{j\geq 0}b_{j-1}\lambda^{-2j},
\end{array}
\end{equation}
we obtain the
recursion equations
\begin{equation}
\begin{array}{rcl}
KG_{j-1}=JG_j,\quad JG_{-1}=0,\ \ j\geq0,
\end{array}
\end{equation}
with $G_j=(a_j,b_j)^T$. Since equation $JG_{-1}=0$ has the general
solution
\begin{equation}
\begin{array}{rcl}
G_{-1}=\alpha_0g_{-1}+\beta_0\hat g_{-1},
\end{array}
\end{equation}
 $G_j$ can be expressed as
\begin{equation}
\begin{array}{lcl}
&&G_j=\alpha_0g_j+\beta_0\hat g_j+\cdots+\alpha_jg_0+\beta_j\hat
g_0\\&& \quad\quad\quad+\alpha_{j+1}g_{-1} +\beta_{j+1}\hat g_{-1},\quad j\geq 0,
\end{array}
\end{equation}
where $\alpha_j$ and $\beta_j$ are arbitrary constants.
Let $\psi$ satisfy the spectral problem (2.1) and an auxiliary
problem
\begin{equation}
\begin{array}{rcl}
\psi_{t_r}=\widetilde{V}^{(r)}\psi,\quad
\widetilde{V}^{(r)}=\lambda\left(\begin{matrix}\widetilde{V}_{11}^{(r)}&
\widetilde{V}_{12}^{(r)}& \widetilde{V}_{13}^{(r)}\cr
\widetilde{V}_{21}^{(r)}&\widetilde{V}_{22}^{(r)}&\widetilde{V}_{23}^{(r)}\cr
\widetilde{V}_{31}^{(r)}&
\widetilde{V}_{32}^{(r)}&\widetilde{V}_{33}^{(r)}\end{matrix}\right),
\end{array}
\end{equation}
where
\begin{equation}
\begin{array}{rcl}
&&\widetilde{V}_{11}^{(r)}=-\frac{1}{3}\partial^2\tilde{a}^{(r)}-\lambda\partial
\tilde{b}^{(r)},\quad
 \widetilde{V}_{12}^{(r)}=\partial \tilde{a}^{(r)}+\lambda \tilde{b}^{(r)},\quad \widetilde{V}_{13}^{(r)}=-2\tilde{a}^{(r)},\\
&&\widetilde{V}_{21}^{(r)}=-\frac{1}{3}\partial^3\tilde{a}^{(r)}-\lambda\partial^2
\tilde{b}^{(r)}-2\lambda v\tilde{a}^{(r)},\quad
\widetilde{V}_{22}^{(r)}=\frac{2}{3}\partial^2\tilde{a}^{(r)},\quad
\widetilde{V}_{23}^{(r)}=- \partial \tilde{a}^{(r)}+\lambda \tilde{b}^{(r)}, \\
&&\widetilde{V}_{31}^{(r)}=-\frac{1}{3}\partial^4\tilde{a}^{(r)}+\lambda^2v\tilde{b}^{(r)},
\ \
\widetilde{V}_{32}^{(r)}=\frac{1}{3}\partial^3\tilde{a}^{(r)}-\lambda\partial^2
\tilde{b}^{(r)}-2\lambda v\tilde{a}^{(r)},\\
&&\widetilde{V}_{33}^{(r)}=-\frac{1}{3}\partial^2\tilde{a}^{(r)}+\lambda\partial
\tilde{b}^{(r)},\ \
\tilde{a}^{(r)}=\sum\limits^r_{j=0}\tilde{a}_{j-1}\lambda^{2(r-j)},\quad
\tilde{b}^{(r)}=\sum\limits^r_{j=0}\tilde{b}_{j-1}\lambda^{2(r-j)},\\
&&\widetilde{G}_j=(\tilde{a}_{j},
\tilde{b}_{j})^T=\tilde{\alpha}_0g_j+\tilde{\beta}_0\hat
g_j+\cdots+\tilde{\alpha}_jg_0+\tilde{\beta}_j\hat
g_0+\tilde{\alpha}_{j+1}g_{-1} +\tilde{\beta}_{j+1}\hat g_{-1},\quad
j\geq -1, \end{array}
\end{equation}
 and the constants $\tilde{\alpha}_{j},\tilde{\beta}_{j}$
are independent of $\alpha_{j}, \beta_{j}.$
 Then the compatibility condition of (2.1) and (2.12) yields
the zero-curvature equation,
 $U_{t_r}-\widetilde{V}_x^{(r)}+[U,\widetilde{V}^{(r)}]=0$, which is equivalent to the hierarchy of nonlinear evolution
 equations
\begin{equation}
\begin{array}{rcl}
v_{t_r}=-\frac{1}{3}\partial^5\tilde{a}_{r-1}=-(\partial
v+2v\partial)\tilde{b}_{r}.
\end{array}
\end{equation}
The first nontrivial member in the hierarchy (2.14) is
\begin{equation}
\begin{array}{rcl}
 &&v_{t_0}=-\frac{1}{3}\partial^5\{\tilde{\alpha}_0v^{-\frac{2}{3}}+\tilde{\beta}_0[-\frac{1}{3}v^{-1}(v^{-\frac{1}{3}})_{xx}+\frac{1}{6}v^{-\frac{2}{3}}({v^{-\frac{1}{3}})_x}^2]\},\\
\end{array}
\end{equation}
which is just the Dym-type equation (1.2) for $\tilde{\alpha}_0=1,\tilde{\beta}_0=0,t_0=t.$

\section{The Baker-Akhiezer Function}\setcounter{equation}{0}

In this section, we shall introduce the vector-valued Baker-Akhiezer function, meromorphic function and trigonal curve associated with
the Dym-type hierarchy. Then we derive a system of
Dubrovin-type differential equations.

We introduce the vector-valued Baker-Akhiezer function
\begin{equation}
\begin{array}{rcl}
&&\psi_x(P,x,x_0,t_r,t_{0,r})=U(v,\lambda)\psi(P,x,x_0,t_r,t_{0,r}),\\
&&\psi_{t_r}(P,x,x_0,t,t_{0,r})=\widetilde{V}^{(r)}(v,\lambda)\psi(P,x,x_0,t_r,t_{0,r}),\\
&&\lambda^{-1}V^{(n)}(v,\lambda)\psi(P,x,x_0,t,t_{0,r})=y(P)\psi(P,x,x_0,t_r,t_{0,r}),\\
&&\psi_1(P,x_0,x_0,t_{0,r},t_{0,r})=1,\quad P=(\lambda,y),\ x, t_r\in\mathbb{C}.
\end{array}
\end{equation}
Here $V^{(n)}=\lambda \left(\begin{matrix}V_{ij}^{(n)}\end{matrix}\right)_{3\times3}$ and
$$\begin{array}{rcl}
&&V^{(n)}_{11}=-\frac{1}{3}\partial^2a^{(n)}-\lambda\partial
b^{(n)},\quad
 V^{(n)}_{12}=\partial a^{(n)}+\lambda b^{(n)},\quad V^{(n)}_{13}=-2 a^{(n)},\\
&&V^{(n)}_{21}=-\frac{1}{3}\partial^3a^{(n)}-\lambda\partial^2
b^{(n)}-2\lambda va^{(n)},\quad
V^{(n)}_{22}=\frac{2}{3}\partial^2a^{(n)},\quad
V^{(n)}_{23}=- \partial a^{(n)}+\lambda b^{(n)}, \\
&&V^{(n)}_{31}=-\frac{1}{3}\partial^4a^{(n)}+\lambda^2vb^{(n)}, \ \
V^{(n)}_{32}=\frac{1}{3}\partial^3a^{(n)}-\lambda\partial^2
b^{(n)}-2\lambda va^{(n)},\\
&&V^{(n)}_{33}=-\frac{1}{3}\partial^2a^{(n)}+\lambda\partial
b^{(n)},\ \
a^{(n)}=\sum\limits_{j=0}^{n}a_{j-1}\lambda^{2(n-j)},\quad
b^{(n)}=\sum\limits_{j=0}^{n}b_{j-1}\lambda^{2(n-j)},
\end{array}$$
in which $a_{j}, b_{j}$  are determined by (2.11). The compatibility
conditions of the first three equations in (3.1) yield that
\begin{equation}
\begin{array}{lcl}
U_{t_r}-\widetilde{V}^{(r)}_x+[U,\widetilde{V}^{(r)}]=0,
\end{array}
\end{equation}
\begin{equation}
\begin{array}{lcl}
-V^{(n)}_x+[U,V^{(n)}]=0,
\end{array}
\end{equation}
\begin{equation}
\begin{array}{lcl}
-V^{(n)}_{t_r}+[\widetilde{V}^{(r)},V^{(n)}]=0.
\end{array}
\end{equation}
A direct calculation shows that $yI-\lambda^{-1}V^{(n)}$ satisfies
(3.3) and (3.4). Hence the characteristic polynomial of Lax matrix
$\lambda^{-1}V^{(n)}$ for the Dym-type hierarchy is a
constant independent of variables $x$ and $t_r$, and possesses following expansion
\begin{equation}
\begin{array}{rcl}
{\mathcal{F}}_m(\lambda,y)={\rm det}(y I-\lambda^{-1}V^{(n)})=y^3+yS_{m}(\lambda)-T_m(\lambda),
\end{array}
\end{equation}
where  $S_{m}(\lambda)$ and $T_m(\lambda)$ are polynomials of $\lambda$ with
constant coefficients
\begin{equation}
\begin{array}{rcl}
&&S_{m}(\lambda)=
 \left|\begin{matrix}V^{(n)}_{11}& V^{(n)}_{12}\cr
                     V^{(n)}_{21}&V^{(n)}_{22}\cr\end{matrix}\right|
 +\left|\begin{matrix}V^{(n)}_{11}& V^{(n)}_{13}\cr
                     V^{(n)}_{31}&V^{(n)}_{33}\cr\end{matrix}\right|
 +\left|\begin{matrix}V^{(n)}_{22}&V^{(n)}_{23}\cr
                 V^{(n)}_{32}&V^{(n)}_{33}\cr\end{matrix}\right|,
\\\\

&&T_m(\lambda)=\left|\begin{matrix}
  V^{(n)}_{11}&V^{(n)}_{12}&V^{(n)}_{13}\cr
  V^{(n)}_{21}&V^{(n)}_{22}&V^{(n)}_{23}\cr
  V^{(n)}_{31}& V^{(n)}_{32}&V^{(n)}_{33}\cr\end{matrix}\right|
  =\left\{\begin{array}{lcl}
\beta_0^3\lambda^{6n+4}+\cdots, \ \  \beta_0\neq0,\alpha_0\in\mathbb{R},\\
-8\alpha_0^3\lambda^{6n+2}+\cdots,\ \ \beta_0=0,\alpha_0\neq0.
\end{array}\right.\\
\end{array}
\end{equation}
It is evident that $T_m(\lambda)$ is a polynomial of degree
$6n+4=3(2n+1)+1$ and $6n+2=3(2n)+2$ as
$\beta_0\neq0$, $\alpha_0\in\mathbb{R}$ and
$\beta_0=0,\alpha_0\neq0$, respectively. Then ${\mathcal{F}}_m(\lambda,y)=0$
naturally leads to a trigonal curve
\begin{equation}
\begin{array}{rcl}
{\mathcal K}_{m-1}: \quad{\mathcal{F}}_m(\lambda,y)=y^3+yS_{m}(\lambda)
-T_m(\lambda)=0,
\end{array}
\end{equation}
with $m=6n+4$ or $m=6n+2$.

For the convenience, we denote the compactification of the curve
 ${\mathcal K}_{m-1}$ by the same symbol ${\mathcal K}_{m-1}$. Thus ${\mathcal
 K}_{m-1}$ becomes a three-sheeted Riemann
 surface of arithmetic genus $m-1$ if it is nonsingular or smooth, which means that
 $\left.(\frac{\partial{\mathcal{F}}_m(\lambda,y)}{\partial\lambda},
 \frac{\partial{\mathcal{F}}_m(\lambda,y)}{\partial y})\right.|_{(\lambda,y)=(\lambda^{\prime},y^{\prime})}\neq(0,0)$ at each
 point $P^{\prime}=(\lambda^{\prime},y^{\prime})\in{\mathcal K}_{m-1}$.

A meromorphic function $\phi(P,x,t_r)$ on ${\mathcal K}_{m-1}$ is defined as
\begin{equation}\label{eqn: 3.8}
\begin{array}{rcl}
\phi(P)=\phi(P,x,t_r)=v^{-\frac{1}{3}}\displaystyle\frac{\psi_{1,x}(P,x,x_0,t_r,t_{0,r})}{\psi_1(P,x,x_0,t_r,t_{0,r})}=v^{-\frac{1}{3}}\displaystyle\frac{\psi_2(P,x,x_0,t_r,t_{0,r})}{\psi_1(P,x,x_0,t_r,t_{0,r})},\quad
P\in{\mathcal K}_{m-1}.
\end{array}
\end{equation}
It infers from (3.1) and (3.8) that
\begin{equation}
\begin{array}{lcl}
 \phi(P)=v^{-\frac{1}{3}}\frac{\displaystyle yV^{(n)}_{23}+C_{m}}{\displaystyle yV^{(n)}_{13}+A_{m}}
     =\frac{\displaystyle v^{-\frac{1}{3}} F_{m}}{\displaystyle y^2V^{(n)}_{23}-yC_{m}+D_{m}}
     =\frac{\displaystyle y^2V^{(n)}_{13}-yA_{m}+B_{m}}{\displaystyle v^{\frac{1}{3}} E_{m-1}},
\end{array}
\end{equation}
where
\begin{equation}
\begin{array}{lcl}
A_m=V^{(n)}_{12}V^{(n)}_{23}-V^{(n)}_{13}V^{(n)}_{22},\ \
C_m=V^{(n)}_{13}V^{(n)}_{21}-V^{(n)}_{11}V^{(n)}_{23},\\
B_m=V^{(n)}_{13}(V^{(n)}_{11}V^{(n)}_{33}-V^{(n)}_{13}V^{(n)}_{31})+V^{(n)}_{12}(V^{(n)}_{11}V^{(n)}_{23}-V^{(n)}_{13}V^{(n)}_{21}),\\
D_m=V^{(n)}_{23}(V^{(n)}_{22}V^{(n)}_{33}-V^{(n)}_{23}V^{(n)}_{32})+V^{(n)}_{21}(V^{(n)}_{13}V^{(n)}_{22}-V^{(n)}_{12}V^{(n)}_{23}),\\
E_{m-1}=(V^{(n)}_{13})^2V^{(n)}_{32}+V^{(n)}_{12}V^{(n)}_{13}(V^{(n)}_{22}-V^{(n)}_{33})-(V^{(n)}_{12})^2V^{(n)}_{23},\\
F_{m}=(V^{(n)}_{23})^2V^{(n)}_{31}+V^{(n)}_{21}V^{(n)}_{23}(V^{(n)}_{11}-V^{(n)}_{33})-V^{(n)}_{13}(V^{(n)}_{21})^2.\\
\end{array}
\end{equation}
 Taking (3.7) and (3.9) into account, we arrive at some important identities among polynomials
$A_{m},B_{m},C_{m}, D_{m},$ $ E_{m-1}, F_{m}, S_m, T_m$:
\begin{equation}
\begin{array}{lcl}
V^{(n)}_{13}F_{m}=V^{(n)}_{23}D_{m}-(V^{(n)}_{23})^2S_{m}-C^2_{m},\\
A_{m}F_{m}=(V^{(n)}_{23})^2T_m+C_{m}D_{m},
\end{array}
\end{equation}
\begin{equation}
\begin{array}{lcl}
V^{(n)}_{23}E_{m-1}=V^{(n)}_{13}B_{m}-(V^{(n)}_{13})^2S_{m}-A_{m}^2,\\
C_{m}E_{m-1}=(V^{(n)}_{13})^2T_m+A_{m}B_{m},
\end{array}
\end{equation}
\begin{equation}
\begin{array}{lcl}
 V^{(n)}_{23}B_{m}+V^{(n)}_{13}D_{m}-V^{(n)}_{13}V^{(n)}_{23}S_{m}+A_{m}C_{m}=0,\\
 V^{(n)}_{13}V^{(n)}_{23}T_m+V^{(n)}_{23}A_{m}S_{m}+V^{(n)}_{13}C_{m}S_{m}-B_{m}C_{m}-A_{m}D_{m}=0,\\
 V^{(n)}_{23}A_{m}T_m+V^{(n)}_{13}C_{m}T_m-B_{m}D_{m}+E_{m-1}F_{m}=0,
\end{array}
\end{equation}
\begin{equation}
\begin{array}{lcl}
E_{m-1,x}=-2V^{(n)}_{13}S_{m}+3B_{m},\\
V^{(n)}_{23}F_{m,x}=-3V^{(n)}_{22}F_{m}+V^{(n)}_{21}(2V^{(n)}_{23}S_{m}-3D_{m}).
\end{array}
\end{equation}
Now, we define the holomorphic mapping $\ast$, changing sheets, by
\begin{equation}\label{eqn: 3.22}
\begin{array}{lcl}
\ast: \left \{\begin{array}{lll}
       {\mathcal K}_{m-1}\rightarrow {\mathcal K}_{m-1}\\
       P=(\lambda, y_i(\lambda))\rightarrow P^\ast=(\lambda,y_{i+1(mod 3)}(\lambda)), \quad i=0,1,2
       \end{array}, \right. \\
P^{\ast\ast} := (P^\ast)^\ast, \quad etc.,
\end{array}
\end{equation}
where $y_i(\lambda), i=0,1,2$, denote the three branches of $y(P)$
satisfying ${\mathcal F}_m(\lambda,y)=0,$ namely,
 \begin{equation}\label{eqn: 3.23}
(y-y_0(\lambda))(y-y_1(\lambda))(y-y_2(\lambda))=y^3+yS_{m}(\lambda)-T_m(\lambda)=0.
\end{equation}
Consequently, we have
\begin{equation}\label{eqn: 3.24}
\begin{array}{lcl}
y_0+y_1+y_2=0,\ \
y_0y_1+y_0y_2+y_1y_2=S_{m}(\lambda),\\
y_0y_1y_2=T_m(\lambda),\ \
y_0^2+y_1^2+y_2^2=-2S_{m}(\lambda),\\
y_0^3+y_1^3+y_2^3=3T_{m}(\lambda),\ \
y_0^2y_1^2+y_0^2y_2^2+y_1^2y_2^2=S^2_{m}(\lambda).
\end{array}
\end{equation}
In what follows, we shall summarize some properties of the meromorphic function $\phi(P,x,t_r)$ without proofs.
\begin{equation}
\begin{array}{lll}
[v^{\frac{1}{3}}\phi(P)]_{xx}+3v^{\frac{1}{3}}\phi(P)[v^{\frac{1}{3}}\phi(P)]_{x}
+v\phi^3(P)=\lambda
v,
 \end{array}
\end{equation}
\begin{equation}\begin{array}{lcl}
[v^{\frac{1}{3}}\phi(P)]_{t_r}=\lambda\Big(\widetilde{V}^{(r)}_{11}+v^{\frac{1}{3}}\widetilde{V}^{(r)}_{12}\phi(P)
+\widetilde{V}^{(r)}_{13}[(v^{\frac{1}{3}}\phi(P))_{x}+v^{\frac{2}{3}}\phi^2(P)]\Big)_x,
\end{array}
\end{equation}
\begin{equation}
v\phi(P)\phi(P^\ast)\phi(P^{\ast\ast})=-\frac{F_{m}}{E_{m-1}},
\end{equation}
\begin{equation}
v^{\frac{1}{3}}[ \phi(P)+\phi(P^\ast)+\phi(P^{\ast\ast})]=\frac{E_{m-1,x}}{E_{m-1}},
\end{equation}
\begin{equation}
\begin{array}{lcl}
 \displaystyle\frac{1}{\phi(P)}+\frac{1}{\phi(P^\ast)}+\frac{1}{\phi(P^{\ast\ast})}=v^{\frac{1}{3}}\Big[-3\frac{V_{22}^{(n)}}
   {V_{21}^{(n)}}-\frac{V_{23}^{(n)}F_{m,x}}{V_{21}^{(n)}F_{m}}\Big],
\end{array}
\end{equation}
\begin{equation}
\begin{array}{lll}
[v^{\frac{1}{3}}\phi(P)+v^{\frac{1}{3}}\phi(P^\ast)+v^{\frac{1}{3}}\phi(P^{\ast\ast})]_{x}+v^{\frac{2}{3}}[\phi^2(P)+\phi^2(P^\ast)\\
\quad\quad+\phi^2(P^{\ast\ast})]
=\displaystyle
-3\frac{V^{(n)}_{11}}{V^{(n)}_{13}}-\frac{V^{(n)}_{12}E_{m-1,x}}{V^{(n)}_{13}E_{m-1}}.
\end{array}
\end{equation}

 {\bf Lemma 3.1.}
Assume that (3.1) and (3.2) hold, and let
$(\lambda,x,t_r)\in{\mathbb{C}}^3$. Then
\begin{equation}
\begin{array}{rrr}
 E_{m-1,t_r}=\lambda E_{m-1,x}[\widetilde{V}^{(r)}_{12}-\frac{\displaystyle\widetilde{V}^{(r)}_{13}}
             {\displaystyle V^{(n)}_{13}}V^{(n)}_{12}]+3\lambda E_{m-1}[\widetilde{V}^{(r)}_{11}
             -\frac{\displaystyle\widetilde{V}^{(r)}_{13}}{\displaystyle V^{(n)}_{13}}V^{(n)}_{11}],
\end{array}\end{equation}
 \begin{equation}
\begin{array}{lcl}
F_{m,t_r}=\lambda
F_{m,x}[\widetilde{V}^{(r)}_{23}-\frac{\displaystyle\widetilde{V}^{(r)}_{21}}
          {\displaystyle V^{(n)}_{21}}V^{(n)}_{23}]+3\lambda F_{m}[\widetilde{V}^{(r)}_{22}
          -\frac{\displaystyle\widetilde{V}^{(r)}_{21}}{\displaystyle
          V^{(n)}_{21}}V^{(n)}_{22}].\\
\end{array}\end{equation}
{\bf Proof.} Differentiating (3.21) with respect to $t_r$ and using (3.19), (3.21) and (3.23), we have
$$\begin{array}{lcl}
\partial_x\partial_{t_r}(\ln E_{m-1})&=&[v^{\frac{1}{3}}\phi(P)+v^{\frac{1}{3}}\phi(P^\ast)+v^{\frac{1}{3}}\phi(P^{\ast\ast})]_{t_r}\\
      &=&\lambda\Big(3\widetilde{V}^{(r)}_{11}+v^{\frac{1}{3}}\widetilde{V}^{(r)}_{12}[\phi(P)+\phi(P^\ast)+\phi(P^{\ast\ast})]
      \\&&+\widetilde{V}^{(r)}_{13}[(v^{\frac{1}{3}}\phi(P)+v^{\frac{1}{3}}\phi(P^\ast)+v^{\frac{1}{3}}\phi(P^{\ast\ast}))_x
     \\&&+v^{\frac{2}{3}}(\phi^2(P)+\phi^2(P^\ast)+\phi^2(P^{\ast\ast}))]\Big)_x\\
      &=&\lambda \big[\displaystyle (\widetilde{V}^{(r)}_{12}-\frac{\widetilde{V}^{(r)}_{13}}{V_{13}^{(n)}}V_{12}^{(n)})\frac{E_{m-1,x}}{E_{m-1}}
           +3(\widetilde{V}^{(r)}_{11}-\displaystyle\frac{\widetilde{V}^{(r)}_{13}}{V_{13}^{(n)}}V_{11}^{(n)})\big]_x.
  \end{array}$$
Integrating the above equation with respect to $x$ and choosing
the integration constant as zero imply the first equation in (3.24). Differentiating (3.20)
with respect to $t_r$, an analogous process shows (3.25).

By observing (2.11) and (3.10), we can easily find that $E_{m-1}$ and
$F_{m}$ are polynomials with respect to $\lambda$ of degree $m-1$
and $m$, respectively. Therefore
\begin{equation}
E_{m-1}(\lambda,x,t_r)=-\epsilon(m)v^{-1}\prod\limits_{j=1}^{m-1}(\lambda-\mu_j(x,t_r)),
\end{equation}
\begin{equation}
F_{m}(\lambda,x,t_r)=\epsilon(m)\prod\limits_{l=0}^{m-1}(\lambda-\nu_l(x,t_r)),
\end{equation}
with $$
\begin{array}{rcl}\epsilon(m)= \left \{\begin{array}{lcl}
\beta_0^3,\ \ m=6n+4,\\
8\alpha_0^3,\ \ m=6n+2.
\end{array}\right.
\end{array}$$
Let us denote
 \begin{equation}
\begin{array}{rcl}
 \hat \mu_j(x,t_r)&=&\Big(\mu_j(x,t_r),y(\hat \mu_j(x,t_r))\Big)\\&=&\Big(\mu_j(x,t_r),-\frac{\displaystyle A_{m}(\mu_j(x,t_r),x,t_r)}
              {\displaystyle V^{(n)}_{13}(\mu_j(x,t_r),x,t_r)}\Big)\in{\mathcal K}_{m-1},\ \  1\leq j\leq m-1,
\end{array}
\end{equation}
\begin{equation}
\begin{array}{rcl}
 \hat \nu_l(x,t_r)&=&\Big(\nu_l(x,t_r),y(\hat \nu_l(x,t_r))\Big)\\&=&\Big(\nu_l(x,t_r),-\frac{\displaystyle C_{m}(\nu_l(x,t_r),x,t_r)}
            {\displaystyle V^{(n)}_{23}(\nu_l(x,t_r),x,t_r)}\Big)\in{\mathcal K}_{m-1},\ \  0\leq l\leq m-1,
\end{array}
\end{equation}
then it is easy to see that the following Lemma holds.

 {\bf Lemma 3.2.}
Suppose the zeros $\{\mu_j(x,t_r)\}_{j=1}^{m-1}$
and $\{\nu_l(x,t_r)\}_{l=0}^{m-1}$ of $E_{m-1}(\lambda,x,t_r)$ and
$F_{m}(\lambda,x,t_r)$ remain distinct for $(x,t_r)\in\Omega_\mu$
and $(x,t_r)\in\Omega_\nu$, respectively, where
 $\Omega_\mu,\Omega_\nu\subseteq\mathbb{C}^2$ are open and connected. Then
 $\{\mu_j(x,t_r)\}_{j=1}^{m-1}$ and $\{\nu_l(x,t_r)\}_{l=0}^{m-1}$ satisfy the Dubrovin-type equations
\begin{equation}
\begin{array}{lcl}
\mu_{j, x}=\frac{\displaystyle
vV^{(n)}_{13}(\mu_j,x,t_r)[3y^2(\hat
\mu_j)+S_{m}(\mu_j)]} {\displaystyle \epsilon(m)
\prod_{\substack{k=1\\k\neq j}}^{m-1}(\mu_j-\mu_k)},
\quad 1\leq j\leq m-1,
\end{array}
\end{equation}
\begin{equation}
\begin{array}{lcl}
&&\mu_{j, t_r}=v\mu_j[V^{(n)}_{13}(\lambda,x,t_r)\widetilde{V}^{(r)}_{12}(\lambda,x,t_r)-\widetilde{V}^{(r)}_{13}(\lambda,x,t_r)V^{(n)}_{12}(\lambda,x,t_r)]_{\lambda=\mu_j}\\
&& \quad\quad\quad  \times\frac{\displaystyle [3y^2(\hat
\mu_j)+S_{m}(\mu_j)]}
{\displaystyle\epsilon(m)\prod_{\substack{k=1\\k\neq
j}}^{m-1}(\mu_j-\mu_k)}, \quad 1\leq j\leq m-1,
\end{array}
\end{equation}
\begin{equation}
\begin{array}{lcl}
&&\nu_{l, x}=\frac{\displaystyle
V^{(n)}_{21}(\nu_l,x,t_r)[3y^2(\hat
\nu_l)+S_{m}(\nu_l)]}
{\displaystyle\epsilon(m)\prod_{\substack{k=0\\k\neq
l}}^{m-1}(\nu_l-\nu_k)}, \quad 0\leq l\leq m-1,
\end{array}
\end{equation}
\begin{equation}
\begin{array}{lcl}
&&\nu_{l, t_r}=\nu_l[V^{(n)}_{21}(\lambda,x,t_r)\widetilde{V}^{(r)}_{23}(\lambda,x,t_r)-\widetilde{V}^{(r)}_{21}(\lambda,x,t_r)V^{(n)}_{23}(\lambda,x,t_r)]_{\lambda=\nu_l}\\
&& \quad\quad\quad \times\frac{\displaystyle [3y^2(\hat
\nu_l)+S_{m}(\nu_l)]}
{\displaystyle\epsilon(m)\prod_{\substack{k=0\\k\neq
l}}^{m-1}(\nu_l-\nu_k)}, \quad 0\leq l\leq m-1.
\end{array}
\end{equation}
{\bf Proof.}  We just need to prove (3.30) for the proofs of (3.31)-(3.33) are similar to (3.30). Substituting $\lambda=\mu_j$ into the first
expression in (3.14), and using (3.12) and (3.29), we get
\begin{equation}
\begin{array}{lcl}
E_{m-1,x}(\mu_j,x,t_r)=V^{(n)}_{13}(\mu_j,x,t_r)[3y^2(\hat
\mu_j)+S_{m}(\mu_j)].
\end{array}
\end{equation}
On the other hand, differentiating (3.26) with respect to $x$ and inserting $\lambda=\mu_j$ into it give
rise to
\begin{equation}
\begin{array}{lcl}
E_{m-1,x}(\mu_j,x,t_r)=\epsilon(m)v^{-1}\mu_{j,
x}\prod_{\substack{k=1\\k\neq
j}}^{m-1}(\mu_j-\mu_k).
\end{array}
\end{equation}
A comparison of (3.34) and (3.35) yields (3.30).

\section{Algebro-geometric Constructions to the Dym-type Hierarchy}\setcounter{equation}{0}

In this section, we shall derive  explicit Riemann theta function
representations for the meromorphic function $\phi(P,x,t_r)$, and in
particular, that of potential $v(x,t_r)$ for the entire Dym-type
hierarchy.

 Taking
the local coordinate $\zeta=\lambda^{-\frac{1}{3}}$ near $P_\infty
\in {\mathcal K}_{m-1}$ in (3.18), the Laurent series of $\phi(P,x,t_r)$
can be explicitly expressed as
\begin{equation}
\begin{array}{rcl}
\phi(P,x,t_r)\underset{\zeta\rightarrow0}=\frac{1}{\zeta}\sum\limits_{j=0}^\infty\kappa_j(x,t_r)\zeta^j,
\quad \quad P\rightarrow P_\infty,
\end{array}
\end{equation}
where
\begin{equation}
\begin{array}{lcl}
 &&\kappa_0=1,\quad \kappa_1=(v^{-\frac{1}{3}})_x, \quad
 \kappa_2=\frac{2}{9}v^{-\frac{5}{3}}v_{xx}-\frac{7}{27}v^{-\frac{8}{3}}v_x^2,\quad
 \kappa_3=\frac{1}{3}v^{-\frac{2}{3}}(v^{-\frac{1}{3}})_{xxx},\\
  &&\displaystyle\kappa_{j}=-\frac{1}{3}[v^{-1}(v^{\frac{1}{3}}\kappa_{j-2})_{xx}+3v^{-\frac{2}{3}}\sum\limits_{i=0}^{j-1}\kappa_{j-1-i}(v^{\frac{1}{3}}\kappa_{i})_x
            +\sum\limits_{i=1}^{j-1}\kappa_{i}\kappa_{j-i}
            \\&&\quad \quad \quad +\sum\limits_{i=1}^{j-1}\sum\limits_{l=0}^{j-i}
                   \kappa_{i}\kappa_{l}\kappa_{j-i-l}], \ \  (j\geq 2).
\end{array}
\end{equation}
Defining the positive divisors on ${\mathcal K}_{m-1}$ of degree $m-1$
$$\begin{array}{lcl}
\mathcal {D}_{P_1,\dots,P_{m-1}}: \left \{\begin{array}{lll}
       {\mathcal K}_{m-1}\rightarrow \mathbb{N}_0,\\
       P \rightarrow \mathcal {D}_{P_1,\dots,P_{m-1}}(P)=
       \left\{\begin{array}{lcl}
       k,\ {\hbox {if}}\ \ P \ \ {\hbox {occurs}}\ \ k\ \ {\hbox {times in}}\ \left\{P_1, \dots, P_{m-1}\right\}\\
       0,\ {\hbox {if}}\ \ P \notin \left\{P_1, \dots, P_{m-1}\right\}
       \end{array}\right.
       \end{array} \right.
\end{array}$$
with $\mathbb{N}_0=\mathbb{N}\cup\{0\}$, one obtains from (3.9) and (4.1) that
the divisor $(\phi(P,x,t_r))$ of $\phi(P,x,t_r)$ is given by
\begin{equation}
 (\phi(P,x,t_r))=\mathcal {D}_{ \hat \nu_0(x,t_r),\hat \nu_1(x,t_r),\dots,\hat \nu_{m-1}(x,t_r)}(P)-\mathcal {D}_{P_\infty,\hat \mu_1(x,t_r),\dots,\hat
 \mu_{m-1}(x,t_r)}(P),
\end{equation}
which implies that $ \hat \nu_0(x,t_r),\hat \nu_1(x,t_r),\dots,\hat
\nu_{m-1}(x,t_r)$ are $m$ zeros and
$P_\infty,\hat \mu_1(x,t_r),$ $\dots,$ $\hat \mu_{m-1}(x,t_r)$ are $m$
poles of $\phi(P,x,t_r)$.

Equip the Riemann surface ${\mathcal K}_{m-1}$ with canonical basis of cycles
$\{\mathfrak{a}_j,\mathfrak{b}_j\}_{j=1}^{m-1}$, which admits intersection numbers
$$\mathfrak{a}_j\circ \mathfrak{b}_k=\delta_{j,k}, \quad  \mathfrak{a}_j\circ \mathfrak{a}_k=0,\quad\mathfrak{b}_j\circ \mathfrak{b}_k=0, \quad j,k=1,\dots,m-1, $$
and the basis of holomorphic differentials
\begin{equation}
\begin{array}{rcl}
&&\widetilde{\omega}_l(P)=\frac{\displaystyle1}{\displaystyle3y^2(P)+S_{m}(\lambda)}\left\{
\begin{array}{lll}
\lambda^{l-1}d\lambda, \ \   1\leq l\leq m-2n-2,\\
y(P)\lambda^{l+2n-m+1}d\lambda, \ \  m-2n-1\leq l\leq m-1,
\end{array} \right. m=6n+4,\\
&&\widetilde{\omega}_l(P)=\frac{\displaystyle1}{\displaystyle3y^2(P)+S_{m}(\lambda)}\left\{
\begin{array}{lcl}
\lambda^{l-1}d\lambda, \ \   1\leq l\leq m-2n-1,\\
y(P)\lambda^{l+2n-m}d\lambda, \ \   m-2n\leq l\leq m-1,
\end{array} \right.
 m=6n+2.
\end{array}
\end{equation}
Thus the period matrices $A$ and $B$ constructed by
\begin{equation}
\begin{array}{rcl}
A_{jk}=\displaystyle\int_{\mathfrak{a}_k}\widetilde{\omega}_j, \quad\quad
B_{jk}=\displaystyle\int_{\mathfrak{b}_k}\widetilde{\omega}_j,
\end{array}
\end{equation}
 are invertible.
Defining the matrix $C=A^{-1}, \tau=CB$, the Riemannian bilinear relation makes it possible to verify that the matrix $\tau$ is symmetric $(\tau_{jk}=\tau_{kj})$ and has positive
definite imaginary part (Im $\tau>0$) (\cite{farkas:1992,griffithsharris:1994}). If we normalize $\widetilde{\omega}=(\widetilde{\omega}_1,\cdots, \widetilde{\omega}_{m-1})$  into new
basis $\omega=(\omega_1,\cdots,\omega_{m-1})$
\begin{equation}
\begin{array}{rcl}
\omega_j=\sum\limits_{l=1}^{m-1}C_{jl}\widetilde{\omega}_l,
\end{array}
\end{equation}
then we have
$$
\displaystyle\int_{\mathfrak{a}_k}\omega_j=\delta_{jk}, \ \
\displaystyle\int_{\mathfrak{b}_k}\omega_j=\tau_{jk},\ \
j,k=1,\dots,m-1.$$
A straightforward calculation yields the following asymptotic expansions:
\begin{equation}
\begin{array}{lcl}
y(P)\underset{\zeta\rightarrow0}=\left\{
\begin{array}{lcl}
\zeta^{-6n-4}[\beta_0-2\alpha_0\zeta^2+O(\zeta^{4})], \ \  P\rightarrow P_\infty, \ \ m=6n+4,\\
-\zeta^{-6n-2}[2\alpha_0+O(\zeta^4)], \ \ \
P\rightarrow P_\infty ,\ \ m=6n+2,
\end{array}
\right.
\end{array}
\end{equation}
\begin{equation}
\begin{array}{lcl}
S_{m}(\lambda)\underset{\zeta\rightarrow0}=\left\{
\begin{array}{lcl}
6\alpha_0\beta_0\zeta^{-12n-6}[1+O(\zeta^{6})], \ \  P\rightarrow P_\infty, \ \ m=6n+4,\\
6\alpha_0\beta_1\zeta^{-12n}[1+O(\zeta^6)],
   \quad P\rightarrow P_\infty,\ \ m=6n+2,
\end{array}
\right.
\end{array}
\end{equation}
\begin{equation}
\begin{array}{rcl}
 \omega_j\underset{\zeta\rightarrow0}=\left\{
\begin{array}{lcl}
(-\displaystyle\frac{C_{j,m-1}}{\beta_0}-\frac{C_{j,4n+2}}{\beta_0^2}\zeta+O(\zeta^3))d\zeta, \quad P\rightarrow P_\infty,\ \  m=6n+4,\\
(-\displaystyle\frac{C_{j,4n+1}}{4\alpha_0^2}+\frac{C_{j,m-1}}{2\alpha_0}\zeta+O(\zeta^3))
d\zeta, \quad P\rightarrow P_\infty, \ \  m=6n+2,
\end{array}
\right.\\ j=1,\dots,m-1.
\end{array}
\end{equation}
Let $\omega_{P_\infty,2}^{(2)}(P)$ denote the normalized Abelian
differential of the second kind, which is  holomorphic on ${\mathcal
 K}_{m-1}\setminus\{P_\infty\}$ with a pole of order 2 at $P_\infty$ and
 satisfies
\begin{equation}
\begin{array}{rcl}
\displaystyle\int_{\mathfrak{a}_j}\omega_{P_\infty,2}^{(2)}(P)=0,
\quad j=1, \dots, m-1.
\end{array}
\end{equation}
\begin{equation}
\begin{array}{rcl}
 \omega_{P_\infty,2}^{(2)}(P)=(\zeta^{-2}+O(1))d\zeta, \  \ \ P \to
 P _{\infty}.
\end{array}
\end{equation}
The $b$-periods of the differential $\omega_{P_\infty,2}^{(2)}$ are
denoted by
\begin{equation}
\begin{array}{rcl}
&&U_2^{(2)}=(U_{2,1}^{(2)},\cdots,U_{2,m-1}^{(2)}), \\
&&U_{2,j}^{(2)}=\displaystyle\frac{1}{2\pi
i}\displaystyle\int_{\mathfrak{b}_j}\omega_{P_\infty,2}^{(2)}(P)=
\left\{
\begin{array}{rcl}
-\displaystyle\frac{C_{j,m-1}}{\beta_0}, \quad m=6n+4,\\
-\displaystyle\frac{C_{j,4n+1}}{4\alpha_0^2}, \quad  m=6n+2,
\end{array}
\right.\quad j=1,\dots,m-1.
\end{array}
\end{equation}
 Furthermore,
let $\omega_{P_\infty,\hat{\nu}_0(x,t_r)}^{(3)}(P)$ denote the
normalized Abelian differential of the third kind defined by
\begin{equation}
\begin{array}{rcl}
 \omega_{P_\infty,\hat{\nu}_0(x,t_r)}^{(3)}(P)=-\displaystyle\frac{y^2(P)+2y^2(\hat{\nu}_0(x,t_r))+S_m(\nu_0(x,t_r))}{\lambda-\nu_0(x,t_r)} \frac{d\lambda}{3y^2(P)+S_{m}(\lambda)}+\sum\limits_{j=1}^{m-1} \gamma_j\omega_j,
\end{array}
\end{equation}
which is holomorphic on ${\mathcal
K}_{m-1}\setminus\{P_\infty,\hat{\nu}_0(x,t_r)\}$ and has simple poles
at $P_\infty$ and $\hat{\nu}_0(x,t_r)$ with corresponding residues $+1$ and $-1$. The constants $\{\gamma_j\}_{j=1}^{m-1}$ are determined by the
normalization condition
\begin{equation}
\begin{array}{rcl}
\displaystyle\int_{\mathfrak{a}_j}\omega_{P_\infty,\hat{\nu}_0(x,t_r)}^{(3)}(P)=0,
\quad j=1, \dots, m-1.
\end{array}
\end{equation}
A direct calculation shows
\begin{equation}
\begin{array}{rcl}
\omega_{P_\infty,\hat{\nu}_0(x,t_r)}^{(3)}(P)\underset{\zeta\rightarrow0}=\left\{
\begin{array}{lll}
(\zeta^{-1}-\delta(m)+O(\zeta))d\zeta,
\quad P\rightarrow P_\infty,\\
(-\zeta^{-1}+O(1))d\zeta,   \quad P\rightarrow
\hat{\nu}_0(x,t_r),
\end{array}
\right.
\end{array}
\end{equation}
with
\begin{equation}
\begin{array}{rcl}\delta(m)= \left\{
\begin{array}{rcl}
\displaystyle\frac{1}{\beta_0}\sum\limits_{j=1}^{m-1} \gamma_j C_{j,m-1},
\ \ \ \ \ \   m=6n+4, \\

\displaystyle\frac{1}{4\alpha_0^2}\sum\limits_{j=1}^{m-1} \gamma_j C_{j,4n+1} ,\quad m=6n+2.
\end{array}
\right.
\end{array}
\end{equation}
 Then
\begin{equation}
\begin{array}{rcl}
\displaystyle\int_{Q_0}^P\omega_{P_\infty,\hat{\nu}_0(x,t_r)}^{(3)}(P)\underset{\zeta\rightarrow0}=
\left\{
\begin{array}{lll}
\ln\zeta+e_{\infty}^{(3)}(Q_0)-\delta(m)\zeta+O(\zeta^2),
\quad P\rightarrow
P_\infty,\\
-\ln\zeta+e_0^{(3)}(Q_0)+O(\zeta),  \quad
P\rightarrow \hat{\nu}_0(x,t_r),
\end{array}
\right.\\
\end{array}
\end{equation}
with $Q_0$ a chosen base point on ${\mathcal
K}_{m-1}\setminus\{P_\infty,\hat{\nu}_0(x,t_r)\}$ and
$e_{\infty}^{(3)}(Q_0),e_0^{(3)}(Q_0)$ two integration constants.

Let ${\mathcal T}_{m-1}=\{\underline{N}+\tau\underline{L},
\underline{N}, \underline{L}\in\mathbb{Z}^{m-1}\}$ be a period lattice. The complex
torus ${\mathcal J}_{m-1}=\mathbb{C}^{m-1}/{\mathcal T}_{m-1}$ is called
a Jacobian variety of ${\mathcal K}_{m-1}$. The Abelian mapping $\mathcal {A}:
{\mathcal K}_{m-1}\rightarrow {\mathcal J}_{m-1}$ is defined as
$$\mathcal {A}(P)=\Big(\mathcal {A}_{1}(P),\cdots,\mathcal
{A}_{m-1}(P)\Big) =\Big(\displaystyle
\int_{Q_0}^P\omega_1,\cdots,\displaystyle
\int_{Q_0}^P\omega_{m-1}\Big) \quad(\hbox{mod} \mathcal {T}_{m-1}),$$
and is extended linearly to the divisor group $\hbox{Div}({\mathcal K}_{m-1})$
$$\mathcal {A}(\sum_{k} n_kP_k)=\sum_{k} n_k\mathcal {A}(P_k),$$
which enables us to give the Abel-Jacobi coordinates for
the nonspecial divisor ${\mathcal
D}_{\hat{\underline{\mu}}(x,t_r)}=\sum^{m-1}\limits_{k=1}\hat{\mu}_k(x,t_r)$
and ${\mathcal
D}_{\hat{\underline{\nu}}(x,t_r)}=\sum^{m-1}\limits_{k=1}\hat{\nu}_k(x,t_r)$:
\begin{equation}
\begin{array}{rcl}
\rho^{(1)}(x,t_r)=\mathcal{A}({\mathcal D}_{\hat{\underline{\mu}}(x,t_r)})=\mathcal
{A}(\sum^{m-1}\limits_{k=1}\hat{\mu}_k(x,t_r))=\sum^{m-1}\limits_{k=1}\mathcal
{A}(\hat{\mu}_k(x,t_r))=\sum^{m-1}\limits_{k=1}
\displaystyle\int_{Q_0}^{\hat{\mu}_k(x,t_r)}\omega,\\
\rho^{(2)}(x,t_r)=\mathcal{A}({\mathcal D}_{\hat{\underline{\nu}}(x,t_r)})=\mathcal
{A}(\sum^{m-1}\limits_{k=1}\hat{\nu}_k(x,t_r))=\sum^{m-1}\limits_{k=1}\mathcal
{A}(\hat{\nu}_k(x,t_r))=\sum^{m-1}\limits_{k=1}
\displaystyle\int_{Q_0}^{\hat{\nu}_k(x,t_r)}\omega,
\end{array}
\end{equation}
where $\underline{\hat{\mu}}(x,t_r)=(\hat{\mu}_1(x,t_r),\cdots,\hat{\mu}_{m-1}(x,t_r)),
\underline{\hat{\nu}}(x,t_r)=(\hat{\nu}_1(x,t_r),\cdots,\hat{\nu}_{m-1}(x,t_r))$.

Let $\theta(\underline{z}(\cdot,\cdot))$ denote the Riemann theta function (\cite{farkas:1992,griffithsharris:1994}) on ${\mathcal K}_{m-1}$. Here $\underline{z}(\cdot,\cdot)$ is defined as
\begin{equation}
\begin{array}{rcl}
&&\underline{z}(P,\underline{\hat{\mu}}(x,t_r))=M-\mathcal {A}(P)+\rho^{(1)}(x,t_r),\quad  P\in{\mathcal K}_{m-1}, \\
&&\underline{z}(P,\underline{\hat{\nu}}(x,t_r))=M-\mathcal {A}(P)+\rho^{(2)}(x,t_r),\quad  P\in{\mathcal K}_{m-1}, \\
\end{array}
\end{equation}
where
 $M$ is the Riemann constant vector.
 Then the Riemann theta function
representation of $\phi(P,x,t_r)$ reads as follows.

{\bf Theorem 4.1.}  Let the curve ${\mathcal K}_{m-1}$ be
nonsingular, $P=(\lambda,y)\in {\mathcal
K}_{m-1}\setminus\{P_\infty\}$, and $(x,t_r),(x_0,t_{0,r}) \in
\Omega_\mu$, where $\Omega_\mu\subseteq\mathbb{C}^2$ is open and
connected. Suppose also that ${\mathcal
D}_{\hat{\underline{\mu}}(x,t_r)}$, or equivalently, ${\mathcal
D}_{\hat{\underline{\nu}}(x,t_r)}$ is nonspecial for
$(x,t_r)\in\Omega_\mu$. Then $\phi(P,x,t_r)$ may be explicitly constructed by the formula
\begin{equation}
\begin{array}{rcl}
\phi(P,x,t_r)=\frac{\displaystyle\theta(\underline{z}(P,\hat{\underline{\nu}}(x,t_r)))
                                   \theta(\underline{z}(P_\infty,\hat{\underline{\mu}}(x,t_r)))}
                     {\displaystyle\theta(\underline{z}(P_\infty,\hat{\underline{\nu}}(x,t_r)))
                                   \theta(\underline{z}(P,\hat{\underline{\mu}}(x,t_r)))}
                                   \exp\Big(e_{\infty}^{(3)}(Q_0)-\displaystyle\int_{Q_0}^P\omega_{P_\infty,\hat{\nu}_0(x,t_r)}^{(3)}\Big).
\end{array}
\end{equation}

{\bf Proof.} Let $\Phi$ denote the right hand side of (4.20), we now have to show $\phi=\Phi$. The Riemann theorem and (4.17) allow us to conclude that $\Phi$ exactly has zeros at the points $\hat{\nu}_0(x,t_r), \hat{\nu}_1(x,t_r)\cdots,\hat{\nu}_{m-1}(x,t_r)$ and poles at $P_{\infty},\hat{\mu}_1(x,t_r),$ $\cdots,\hat{\mu}_{m-1}(x,t_r).$
It infers from (4.17) that
\begin{equation}
\begin{array}{rcl}
\Phi \underset{\zeta\to
0}=\zeta^{-1}[1+O(\zeta)], \ \ P \to P_{\infty},
\end{array}
\end{equation}
which together with (4.1) gives
\begin{equation}
\begin{array}{rcl}
\frac{\Phi}{\phi} \underset{\zeta\to
0}=\frac{\zeta^{-1}[1+O(\zeta)]}{\zeta^{-1}[1+O(\zeta)]}=1+O(\zeta),\ \ P \to P_{\infty}.
\end{array}
\end{equation}
Applying the Riemann-Roch theorem, we get $\frac{\Phi}{\phi}=1$, which completes the proof.

Based on the above results, we will obtain the Riemann theta
function representations of solutions for the entire Dym-type
hierarchy immediately.

{\bf Theorem 4.2. } Assume that the curve $\mathcal {K}_{m-1}$ is
nonsingular and let $(x,t_r)\in\Omega_\mu$, where
$\Omega_\mu\subseteq\mathbb{C}^2$ is open and connected. Suppose
also that $\mathcal {D}_{\underline{\hat{\mu}}(x,t_r)}$, or
equivalently, $\mathcal {D}_{\underline{\hat{\nu}}(x,t_r)}$ is
nonspecial for $(x,t_r)\in\Omega_\mu$. Then the Dym-type
hierarchy admits algebro-geometric solutions
\begin{equation}
\begin{array}{lcl}
(v^{-\frac{1}{3}})_x&=&\partial_{U_2^{(2)}}\ln\displaystyle\frac{\theta(\underline{z}(P_\infty,
           \underline{\hat{\mu}}(x,t_r)))}{\theta(\underline{z}(P_\infty,
           \underline{\hat{\nu}}(x,t_r)))}+\delta(m),
\end{array}
\end{equation}
with $\delta(m)$ defined in (4.16).

{\bf Proof.} From (4.9), (4.12), (4.18) and (4.19) it follows that
$$
\begin{array}{rcl}
\underline{z}(P,
           \underline{\hat{\mu}}(x,t_r))&=&M-\displaystyle\int_{Q_0}^{P}\omega+\sum^{m-1}\limits_{k=1}\displaystyle\int_{Q_0}^{\hat{\mu}_k(x,t_r)}\omega\\
           &=&(\dots,
           M_j-\displaystyle\int_{Q_0}^{P_\infty}\omega_j+\sum^{m-1}\limits_{k=1}\displaystyle\int_{Q_0}^{\hat{\mu}_k(x,t_r)}\omega_j-\displaystyle\int_{P_\infty}^{P}\omega_j,\dots)\\
&\underset{\zeta\to 0}=&(\dots,
           M_j-\displaystyle\int_{Q_0}^{P_\infty}\omega_j+\sum^{m-1}\limits_{k=1}\displaystyle\int_{Q_0}^{\hat{\mu}_k(x,t_r)}\omega_j-U_{2,j}^{(2)}\zeta+O(\zeta^2),\dots), \ \ P  \to P_{\infty}.\\
\end{array}
$$
Hence
\begin{equation}
\begin{array}{rcl}
&&\displaystyle \frac{\theta(\underline{z}(P,
\underline{\hat{\mu}}(x,t_r)))}
           { \theta(\underline{z}(P_\infty,
           \underline{\hat{\mu}}(x,t_r)))}
\underset{\zeta\to 0}=\displaystyle\frac{\theta(\dots,
           M_j-\displaystyle\int_{Q_0}^{P_\infty}\omega_j+\sum^{m-1}\limits_{k=1}\displaystyle\int_{Q_0}^{\hat{\mu}_k(x,t_r)}\omega_j-U_{2,j}^{(2)}\zeta+O(\zeta^2),\dots)}{\theta(\underline{z}(P_\infty,
           \underline{\hat{\mu}}(x,t_r)))} \\

&&\underset{\zeta\to
0}=\displaystyle\frac{\theta(\underline{z}(P_\infty,
           \underline{\hat{\mu}}(x,t_r)))-\left.[\sum\limits_{j=1}^{m-1}U_{2,j}^{(2)}\frac{\partial}{\partial
z_j}\theta(M-\mathcal
{A}(P_\infty)+\rho^{(1)}(x,t_r)-U^{(2)}_2\zeta+O(\zeta^2))]\right|_{\zeta=0}\zeta+O(\zeta^2)}
{\theta(\underline{z}(P_\infty,
           \underline{\hat{\mu}}(x,t_r)))}\\

&&\underset{\zeta\to
0}=1-\displaystyle\frac{\partial_{U_2^{(2)}}\theta(\underline{z}(P_\infty,
           \underline{\hat{\mu}}(x,t_r)))}{\theta(\underline{z}(P_\infty,
           \underline{\hat{\mu}}(x,t_r)))}\zeta+O(\zeta^2)\\

 && \underset{\zeta\to
0}=1-[\partial_{U_2^{(2)}}\ln\theta(\underline{z}(P_\infty,
           \underline{\hat{\mu}}(x,t_r)))]\zeta+O(\zeta^2),        \ \ P \to P_{\infty},
\end{array}
\end{equation}
where $\partial_{U_2^{(2)}}\theta(\underline{z}(P_\infty,
           \underline{\hat{\mu}}(x,t_r)))=\left.[\partial_{U_2^{(2)}}\theta(M-\mathcal
{A}(P_\infty)+\rho^{(1)}(x,t_r)-U^{(2)}_2\zeta+O(\zeta^2))]\right|_{\zeta=0}$
and
$\partial_{U_2^{(2)}}=\sum\limits_{j=1}^{m-1}U_{2,j}^{(2)}\frac{\partial}{\partial
z_j}$. Quite similarly, we get
\begin{equation}
\begin{array}{rcl}
\displaystyle \frac{\theta(\underline{z}(P,
\underline{\hat{\nu}}(x,t_r)))}
           { \theta(\underline{z}(P_\infty,
           \underline{\hat{\nu}}(x,t_r)))}
\underset{\zeta\to
0}=1-[\partial_{U_2^{(2)}}\ln\theta(\underline{z}(P_\infty,
           \underline{\hat{\nu}}(x,t_r)))]\zeta+O(\zeta^2), \ \ \ \ P
\to P_{\infty},
\end{array}
\end{equation}
 with
$\partial_{U_2^{(2)}} \theta(\underline{z}(P_\infty,
           \underline{\hat{\nu}}(x,t_r)))=\left.[\partial_{U_2^{(2)}}\theta(M-\mathcal
{A}(P_\infty)+\rho^{(2)}(x,t_r)-U^{(2)}_2\zeta+O(\zeta^2))]\right|_{\zeta=0}$.

By virtue of (4.17), (4.20), (4.24) and (4.25), we have
\begin{equation}
\begin{array}{rcl}
\phi(P,x,t_r) &\underset{\zeta\to 0}=&
\{1-[\partial_{U_2^{(2)}}\ln\theta(\underline{z}(P_\infty,
           \underline{\hat{\nu}}(x,t_r)))]\zeta+O(\zeta^2)\}\{1+[\partial_{U_2^{(2)}}\ln\theta(\underline{z}(P_\infty,
           \underline{\hat{\mu}}(x,t_r)))]\zeta\\&&+O(\zeta^2)\}
           \times[\zeta^{-1}+\delta(m)+O(\zeta)]\\
& \underset{\zeta\to
0}=&[\zeta^{-1}+\partial_{U_2^{(2)}}\ln\displaystyle\frac{\theta(\underline{z}(P_\infty,
           \underline{\hat{\mu}}(x,t_r)))}{\theta(\underline{z}(P_\infty,
           \underline{\hat{\nu}}(x,t_r)))}+\delta(m)+O(\zeta)],

           \ \ P \to P_{\infty}.
\end{array}
\end{equation}
Comparing (4.1) with (4.26), we arrive at (4.23).

\vspace{0.3cm}
 \noindent\textbf{Acknowledgments}

 This work was supported by National Natural Science Foundation of
China (project nos.11331008, 11401230 and 11501526 ), Cultivation Program for Outstanding Young Scientific talents of the Higher Education Institutions of Fujian Province in 2015, and Promotion Program for Young and Middle-aged Teacher in Science and Technology Research of Huaqiao University (project no. ZQN-PY301).

\end{document}